\documentstyle[11pt,aaspp4]{article}

\begin{document}

\title {\bf The FIRST Bright QSO Survey}

\author{Michael~D.~Gregg}
\affil{Institute for Geophysics and Planetary Physics\\
Lawrence Livermore National Laboratory\\
gregg@igpp.llnl.gov}

\author{Robert H. Becker}
\affil{University of California at Davis\\
and\\
Institute for Geophysics and Planetary Physics\\
Lawrence Livermore National Laboratory\\
bob@igpp.llnl.gov}

\author{Richard L. White}
\affil{Space Telescope Science Institute\\
rlw@stsci.edu}

\author{David J. Helfand\footnote{Visiting Astronomer, Kitt Peak National
Observatory, National Optical Astronomy Observatory}}
\affil{Columbia Astrophysics Laboratory\\
djh@carmen.phys.columbia.edu}

\author{Richard G.~McMahon}
\affil{Institute of Astronomy, Cambridge\\
rgm@ast.cam.ac.uk}

\author{Isobel M. Hook}
\affil{University of California at Berkeley\\
imh@bigz.berkeley.edu}

\begin{abstract}

The FIRST radio survey provides a new resource for constructing a
large quasar sample.  With source positions accurate to better than
1\arcsec\ and a point source sensitivity limit of 1 mJy, it reaches 50
times deeper than previous radio catalogs.  We report here on the
results of the pilot phase for a FIRST Bright Quasar Survey (FBQS).
Based on matching the radio catalog from the initial 300 deg$^2$ of
FIRST coverage with the optical catalog from the Automated Plate
Machine (APM) digitization of Palomar Sky Survey plates, we have
defined a sample of 219 quasar candidates brighter than E = 17.50.  We
have obtained optical spectroscopy for 151 of these and classified 25
others from the literature, yielding 69 quasars or Seyfert~1 galaxies,
of which 51 are new identifications.  The brightest new quasar has an
E magnitude of 14.6 and z = 0.91; four others are brighter than E =
16.  The redshifts range from z=0.12 to 3.42.  Half of the detected
objects are radio quiet with L$_{\rm 21cm} < 10^{32.5}$ ergs/s.  We
use the results of this pilot survey to establish criteria for the
FBQS that will produce a quasar search program which will be 70\%
efficient and 95\% complete to a 21-cm flux density limit of 1.0 mJy.

\end{abstract}

\section {Introduction}

The number of QSOs has ballooned over the past decade, largely as a
result of systematic large-area surveys.  The majority of these
surveys have been based on optical selection criteria, such as the
Large Bright Quasar Survey (LBQS; Hewett, Foltz, \& Chaffee 1995) and
the Edinburgh Quasar Survey (Goldschmidt et al. 1992), in marked
contrast to the earliest radio-selected QSO searches.  The recent
emphasis on optically-selected samples is due in part to their higher
efficiency; previous radio selected samples suffer mainly from the
difficulty of identifying optical counterparts from poor radio
positions.  Additionally, past radio selected samples have been
largely insensitive to radio quiet objects, which make up the
preponderance of the QSO population.  Yet optically selected quasar
samples have their own disadvantages, possibly excluding QSOs with
unusual colors.  For example, it has recently been suggested that
there is a large population of 'dusty' QSOs which has been missed by
recent QSO surveys (Webster et al. 1994).  A radio selected sample
can, in principle, be immune to optical color selection effects.

We show here that the NRAO\footnote{The National Radio Astronomy
Observatory is operated by Associated Universities, Inc., under
cooperative agreement with the National Science Foundation} Very Large
Array (VLA) FIRST\footnote{The FIRST Survey World Wide Web homepage is
http://sundog.stsci.edu} survey (Becker, White and Helfand 1995;
hereafter BWH) is capable of generating a radio selected sample of
quasars which can achieve the very high efficiency of optical surveys.
Such a new complete sample of radio-selected QSOs will help address
several unresolved questions, such as the surface density of bright
quasars and whether or not there is differential evolution between
radio loud and radio quiet QSOs.

We present the results of the pilot phase of the FIRST Bright Quasar
Survey (FBQS) based on the initial 300 deg$^2$ imaged by the VLA FIRST
survey in 1993.  The early results, especially some of the more
unusual quasars in the sample, may be of particular interest to
others.  In $\S$ II we discuss the selection of candidate QSOs based
on a comparison of the FIRST survey catalog of radio sources with the
Automated Plate Machine (APM) catalog of POSS I objects (McMahon \&
Irwin 1992).  In $\S$ III, we describe the optical observations used
to confirm sources as QSOs and present the spectroscopic results.  We
then discuss the efficiency of the survey in $\S$ IV, concluding in
$\S$ V with a discussion of future plans for the survey.

\section {Selection of QSO Candidates}

The preliminary VLA FIRST catalog of radio sources covered 306 deg$^2$
in a narrow strip through the north Galactic pole (BWH) and contained
$\sim 27,000$ sources, complete to a flux density limit of 1~mJy for
point sources.  Comparisons between the FIRST catalog and standard
radio calibration sources indicate that the systematic astrometry
errors are $< 0\farcs2$ in both RA and Dec.  Extensive tests also
indicate that even for the faintest radio sources, positions are
accurate to $\pm 1\arcsec$ (90\% confidence).  Nonetheless, in matching to
optical counterparts we required only 2\arcsec\ agreement in position, in
part to allow an independent check on the radio positional accuracy
(see section IV).

QSO candidates were selected by matching FIRST survey sources to the
APM catalog of the Palomar Observatory Sky Survey (POSS~I).  Objects
classified by the APM as stellar on either of the two POSS~I
emulsions, O (blue) or E (red), brighter than 17.5 magnitude on the E
plate, and within 2\arcsec\ of a FIRST radio source were included in
this pilot quasar survey.  In keeping with the desire to avoid optical
selection effects, no color cut was imposed on this initial candidate
list.

The original National Geographic-Palomar Observatory Sky Survey was
carried out using Eastman 103a-O and 103a-E emulsions.  Plots of the
effective system response for each are given in Minkowski \& Abell
(1963).  The ``O'' and ``E'' passbands have effective wavelengths of
roughly 4200\AA\ and 6400\AA, and effective widths of approximately
1200\AA\ and 400\AA, respectively.  Except for the narrowness of the E
passband, these are similar to Johnson B and Cousins R; the color
transformation for normal stars is (B-R) = 0.875(O-E) + 0.073 (Bessell
1996, private communication).  In what follows, we have chosen not to
transform the magnitudes from O and E to a more standard system for
two reasons.  Objects with emission lines will not transform in a
straightforward manner as do stars.  Also, the errors in O and E are
rather large.  In a comparison of 33 QSOs for which the Automated
Plate Scanner (http://isis.spa.umn.edu) magnitudes were available
on-line, we obtain a scatter of 0.4 magnitudes in both O and E and
0.35 magnitudes in O-E.  Much of the scatter can perhaps be attributed
to the lack of field-by-field calibration of the APM catalog.  Within
this error range, O and E can be considered approximately equal to B
and R magnitudes over the color range of our sample.  We have begun an
observing program at Lick Observatory to improve the optical
photometry of the QSOs in the sample, however, the completeness of the
survey remains problematic because of the large errors.  We will
address this issue in more detail in future installments of this
survey.

The other selection criteria also raise some issues regarding
completeness.  The requirement of a 2\arcsec\ positional coincidence
discriminates against QSOs associated with extended radio sources
(lobes), selecting only those QSOs with nuclear radio emission or
compact morphology as seen by the FIRST survey.  Additional QSOs might
be found by relaxing the coincidence requirements but only at the
expense of additional chance coincidences.  Part of the intent of the
pilot survey is to explore the level of completeness arising from the
morphological considerations.

With the above criteria, 219 QSO candidates were selected, 0.8\% of
the total FIRST catalog.  Ninety-seven were classified as stellar on
both POSS~I emulsions.  Another 29 were classified as stellar on the E
plates only and 93 as stellar on the O plates only.  Since the O
magnitudes are often much fainter than the E magnitudes, their
classification is typically less reliable, especially near the plate
limit where the tendency is for the APM to classify these objects as
stellar.  This leads to a large number of apparently normal, red
galaxies being included in the candidate list.

Using the NASA Extragalactic Database (NED), 25 of the 219 candidates
were found to be already identified in the literature as QSOs or
galaxies, leaving 194 objects for spectroscopic follow-up.

\section {Optical Observations}

The spectroscopy was carried out at Lick Observatory, Kitt Peak
National Observatory,
\footnote{Kitt Peak National Observatory, NOAO, is operated by the
Association of Universities for Research in Astronomy, Inc. (AURA),
under cooperative agreement with the National Science Foundation}
and La Palma.  The observations at Lick Observatory were made on the
Shane 3-m telescope with the Kast spectrograph spanning the wavelength
range 3500-8000\AA.  The Kitt Peak spectra are from the 4-m telescope
with somewhat redder wavelength coverage (4500-9000\AA) and have some
overlap of second order at the red end.  The La Palma 2.5-m telescope
spectra span 5000 - 8000\AA.  All the spectra have a resolution of
$\sim 5$\AA.  Integration times were typically 10-15 minutes.
Observing conditions varied markedly both in transparency and seeing.
Sample QSO spectra are shown in Figure~1, including the brightest new
quasar, BQ~0751+2919 with E magnitude = 14.6, and the highest redshift
new quasar, BQ~0933+2845 with z = 3.42.

To date we have obtained optical spectra for 151 of the 194 previously
unclassified QSO candidates.  Because of the large range in
brightness, the spectra vary in quality from signal-to-noise of 50 to
$\sim 5$.  We have classified objects into five categories.  Spectra
with broad emission lines with $\sigma \gtrsim 1000~{\rm km s^{-1}}$,
much greater than typical galaxy velocity dispersions, are classified
as QSOs.  This classification draws no distinction between bona fide
QSOs and Seyfert~1 galaxies; better optical images are required for
this refinement, although the optical luminosities can be used to make
a rough separation between the two (see below).  Featureless spectra
are designated as BL Lacs.  Spectra with narrow emission lines,
$\sigma \sim 150-300~{\rm km s^{-1}}$, are classed as ELG, making no
distinction among true AGN, starburst, and ordinary star forming
galaxies.  The two remaining categories are galaxies with absorption
lines only (ALG), and Galactic stars.  All but 4 of the spectra
provided positive classifications; these remaining 4 are among the
lowest S/N spectra but are good enough to confidently rule out the
presence of strong emission lines.  On the basis of the radio
association and a relatively red optical spectral energy distribution,
we tentatively classify these 4 as ALGs, though the possibility exists
that they are red Bl Lacs.  The classifications of the 151 spectra and
25 NED-identified objects break down as 69 QSOs, 3 BL Lacs, 32 ELG, 41
ALG, and 31 stars.  Despite the bright magnitude limit, only 15 of the
QSOs were previously cataloged.

For each of the 69 QSOs found in the pilot survey region of 306
deg$^2$, Table 1 lists RA and Dec (J2000), O and E apparent magnitudes
from the APM catalog, 20 cm flux density (S$_{21}$) for the nuclear
radio source, total 20 cm flux density including radio lobes (for
objects with multiple components), emission line redshift, the
logarithm of the inferred radio luminosity for the nuclear radio
source, and the absolute E magnitude.  For consistency with other
studies, we adopt ${\rm H_{o} = 50~km~s^{-1} Mpc^{-1}}$ and ${\rm q_o}
= 0.5$ to compute the distance-dependent quantities using the relations
from Weedman (1986).  We assume that the spectral energy distributions
are described by a simple power law in frequency with an exponent of
$\alpha_{\rm R} = -0.5$.  Notes to Table~1 describe the radio
morphology where it differs from a single compact component.  Eleven
of the QSOs have extended emission on a scale $> 10\arcsec$ while
another five objects have measured single component sizes $>
2\farcs5$, the resolution limit of the FIRST Survey (White et
al. 1996).

In Tables~2 and 3, we list similar data for the narrow emission line
galaxies and the absorption line galaxies respectively.  Data for the
Galactic stars are contained in Table~4.  Many, but not all, of the
stars are chance coincidences (Becker et al. 1995).

Figure 2 shows the distribution of observed nuclear radio flux for the QSOs;
every QSO in the sample with ${\rm S_{21\rm cm}} > 195$ mJy has been
previously identified, an indication of the completeness threshold of
previous radio-selected samples.

About 50\% of the QSOs in the sample are radio quiet with ${{\rm
L_{21cm} < 10^{32.5} erg~s^{-1}~Hz^{-1}}}$, using a simple luminosity
cut as the criterion for radio loudness (Schneider et al.  1992).
With a flux density limit of 1~mJy, some radio quiet QSOs will be
detectable in the FIRST survey out to a redshift of $\sim 2$.  A
histogram of the radio luminosities (Figure 3) has a suggestion of a
deficit at ${\rm L_{21cm} \approx 10^{33} erg~s^{-1}~Hz^{-1}}$,
hinting at a bimodal distribution, roughly consistent with radio loud
and radio quiet objects.  As the survey progresses and the sample
grows, the reality of the bimodality will be resolved, though it is
already apparent that there is considerable overlap in the
distributions.  The narrow emission line galaxy (ELG) and absorption
line galaxy (ALG) luminosities are shown for comparison (dashed line).
The latter two classes have indistinguishable distributions.

Figure 4 displays a histogram of the redshift distribution for the 69
QSOs.  The radio loud QSOs have been shaded to illustrate clearly the
bias against detecting high redshift radio quiet QSOs in a
radio-selected sample.  The fraction of radio loud objects increases
from $\sim 15\%$ for z$< 0.5$ to $\sim 50\%$ for $0.5 \leq {\rm z}
\leq 1.5$ to $\sim 75\%$ for z$ > 1.5$.  The overall distribution is
roughly consistent with that of the LBQS, though detailed comparisons
must await better statistics when we have a larger sample.

Figure 5 shows the histogram of absolute E magnitudes derived from the
APM magnitudes, adopting $\alpha_{\rm opt} = -0.5$ and the same
cosmological assumptions as above.  These have not been corrected for
Galactic extinction or contributions from emission lines; the errors
in the APM magnitudes dominate these corrections.  The QSOs again
suggest a bimodal distribution.  The usually accepted luminosity
cutoff between Seyfert~1 galaxies and QSOs is M$_{\it B} \approx -23$,
which is equivalent to M$_{\it E} \sim -24$ for typical O-E of 1.  The
low luminosity component, containing $\sim18$ objects, is probably the
Seyfert~1 contribution to the sample and perhaps accounts for the
bimodality.  All but five of the 18 were classified as stellar on both
POSS plates, so deeper optical images are necessary to confirm them as
Seyferts.  The ELG and ALG distributions are again very similar and
are plotted for comparison as a single dashed histogram.  Eight of the
nearest absorption line objects have either unusually bright APM
magnitudes, or the APM could not assign a reliable magnitude because
of complicated image structure; these are indicated by a 0.\ in
Table~3.  All of these are previously cataloged objects and a
comparison with data from NED shows that the APM numbers are in error
by many magnitudes.  These have been excluded from the histogram and
no entry for M$_{\rm E}$ is present in Table~3.

\section {Efficiency and Completeness of the Survey}

The results now allow us to evaluate the original selection criteria
and formulate new improved criteria for future observations.  The two
primary selection criteria were the radio/optical position agreement
and the APM stellar/nonstellar classification.  Of the 219 candidates,
97 were classified as stellar on both plates, 29 as stellar on the E
plate only, and 93 as stellar on the blue plate only.  Of the 69
confirmed QSOs, the three comparable numbers are 55, 3 and 11.
Clearly these numbers leave open the possibility that some QSOs are
classified as nonstellar on both plates; we will return to this issue
below.  The reliability of the APM classifier is magnitude dependent:
all 11 QSOs which are stellar on the O plate only are 18th magnitude
or brighter on the O plate.  Seventy-four fainter candidates are
included in the initial sample of 219 because of a stellar
classification from only the O plate; all but 6 of these have O-E
colors redder than 2.0.  Thirty of these have spectroscopic
classifications and are all either ELG or ALG.
In the continuation of the survey these can be eliminated by selecting
objects with O-E $<$ 2.0 (see below).  A color magnitude diagram of
all the survey objects with redshifts and spectral classifications is
shown in Figure 6.

The angular separation selection criterion turned out to be too
generous.  Of the 69 QSOs in the sample, 64 have separations between
the radio position and optical counterpart of $<$ 1\farcs0, 3
between 1\farcs0 - 1\farcs1, and 3 between 1\farcs1-2\farcs0.  The 2
QSOs with the greatest separation are extended radio sources.  Of the
original 219 candidates, over 25\% lie outside 1\farcs1, so a tighter
matching radius will eliminate many false candidates at the expense of
only $\sim 5\%$ of the QSOs.

Although color was not used in the original selection, the survey is
not finding very red quasars: only five QSOs in the sample are redder
than 1.5, the reddest with O-E = 1.74.  This is perhaps because of the
relatively bright magnitude limit; typical highly reddened QSOs will
be much fainter in the optical (Webster et al. 1995) and will not
appear in our candidate list.  Imposing a color cut on our candidate
sample will eliminate a sizable fraction of the non-QSO objects while
sacrificing at most only a tiny number of QSOs, albeit potentially
interesting ones.  In the whole sample, 104 candidates out of 219 are
redder than 1.75.  We classified 63 of these objects
spectroscopically; 34 as galaxies, 18 as ELG, 11 as Galactic stars,
and none as QSOs.  For an additional test of using a color cut in
selecting candidates, we matched the V\'{e}ron-Cetty \& V\'{e}ron
(1996) QSO catalog against the APM catalog.  Of the 380 QSOs that were
within 1\arcsec\ of an APM object brighter than 17.5 on the E plate,
only one was redder than 2.00, confirming the utility of color as a
selection criterion.

A more efficient observing program, then, would restrict candidates to
those optical counterparts classified as stellar on either POSS plate,
falling within a 1\farcs1 matching radius, and with colors bluer than
E-O = 2.0.  If the pilot survey is typical, such a sample would be
70\% QSOs and 95\% complete.  Only two of the 68 candidates still
without spectroscopic classification in the current sample would
survive the new selection criteria.

Two potential causes of incompleteness are the absence of a
core radio source in a radio loud QSO, and a nonstellar classification
on both POSS plates.  To estimate the magnitude of these two effects,
we have carried out several tests.  As mentioned above, the V\'{e}ron
catalog of QSOs was matched with the APM catalog, selecting objects
with E $< 17.5$ and separations $< 1\farcs0$; 380 matches resulted.
Of these 380 QSOs, 27 (7\%) were classified as nonstellar on both POSS
plates; however, only 6 ($< 2\%$) had M$_{\rm B}$ brighter than -24.3,
implying that most of these are Seyfert~1 galaxies 
and are probably correctly classified as
nonstellar.

It is more difficult to estimate the fraction of radio loud QSOs
without core components.  We have inspected the FIRST survey images
around the positions of all the V\'{e}ron QSOs in the survey area and
have found one radio loud QSO (B2 1248+30) with E = 17.5 that
our selection criteria missed because it has no core
radio component.  This suggests that few QSOs are missed because of
this effect, but the test is inconclusive because other
radio-selected QSO surveys may suffer from a similar incompleteness.


\section{Summary and Future Plans}

We have shown that the FIRST radio survey can be used in conjunction
with the POSS plate material as the basis for an efficient QSO survey,
highly complete for radio loud sources.  With the 1994 survey data,
the current FIRST catalog covers a total 1550 deg$^2$ and includes
over 138,000 radio sources.  Using our revised screening criteria, we
have prepared a list of $\sim 400$ additional QSO candidates for
spectroscopic follow-up.  With this larger sample, selected in a
uniform fashion and covering a large contiguous area of sky, we will
begin to be able to address important questions regarding quasar radio
and optical luminosity functions and their evolution, the large scale
distribution of quasars, and the differences between radio loud and
radio quiet objects.

\acknowledgments

We thank Chris Impey for a thorough referee's report and
numerous constructive comments.
The FIRST Survey is supported by grants from the National Science
Foundation, NATO, the National Geographic Society, Sun Microsystems,
and Columbia University.  We thank Michael Strauss for taking some
preliminary data in connection with this project.  Part of the work
reported here was done at the Institute of Geophysics and Planetary
Physics, under the auspices of the U.S. Department of Energy by
Lawrence Livermore National Laboratory under contract
No.~W-7405-Eng-48.  We acknowledge extensive use of the NASA/IPAC
Extragalactic Database (NED) which is operated by the Jet Propulsion
Laboratory, Caltech, under contract with the National Aeronautics and
Space Administration.  This is Contribution Number 594 of the Columbia
Astrophysics Laboratory.

\pagebreak


\voffset -0.5in

\begin{figure}
\plotone{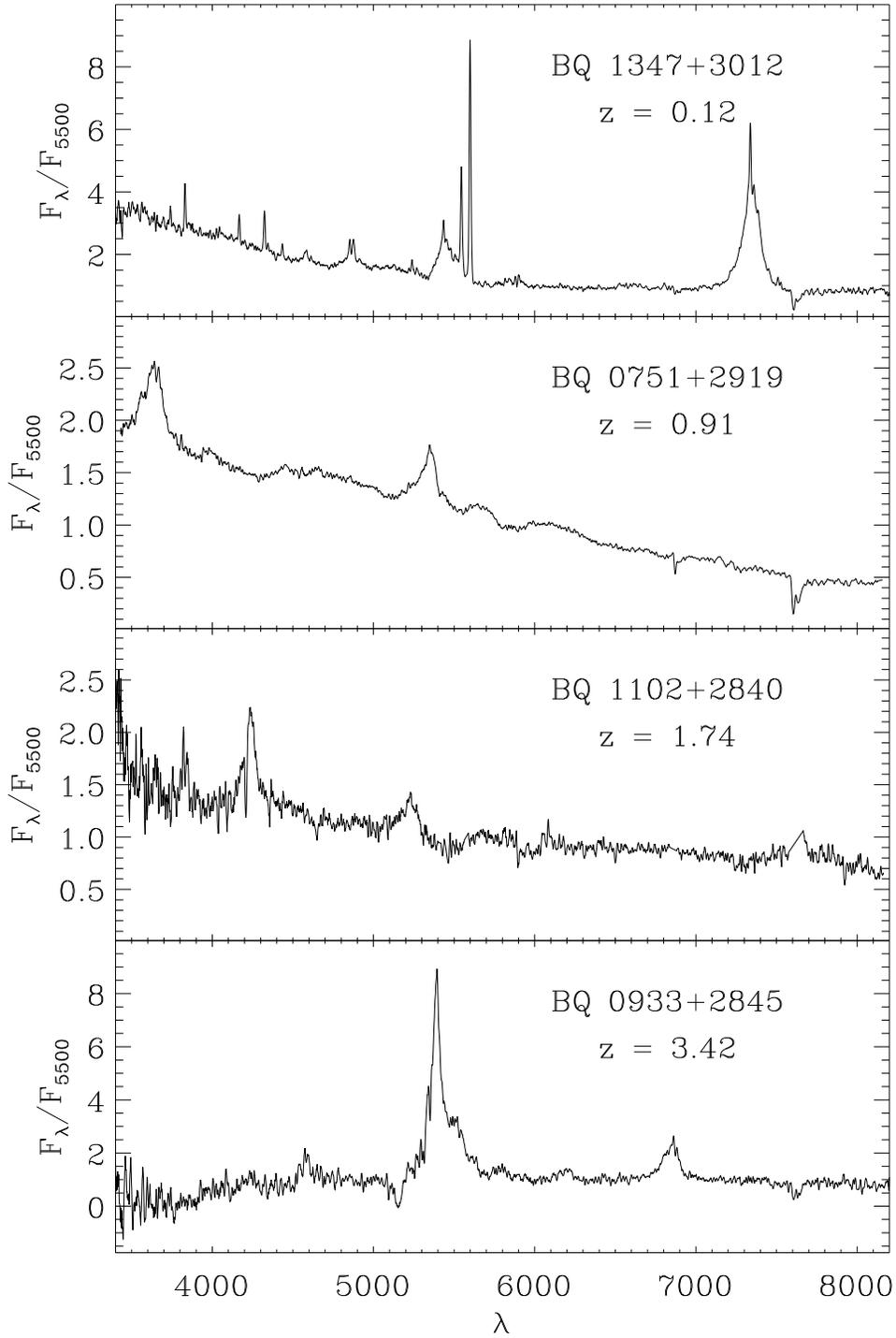}
\caption{Sample optical spectra from the Lick Observatory Kast Double
Spectrograph.  BQ~0751+2919 is the brightest of the new quasars, with E = 14.6;
BQ~0933+2845 is the highest redshift new quasar, with z=3.42.}
\end{figure}

\begin{figure}
\plotone{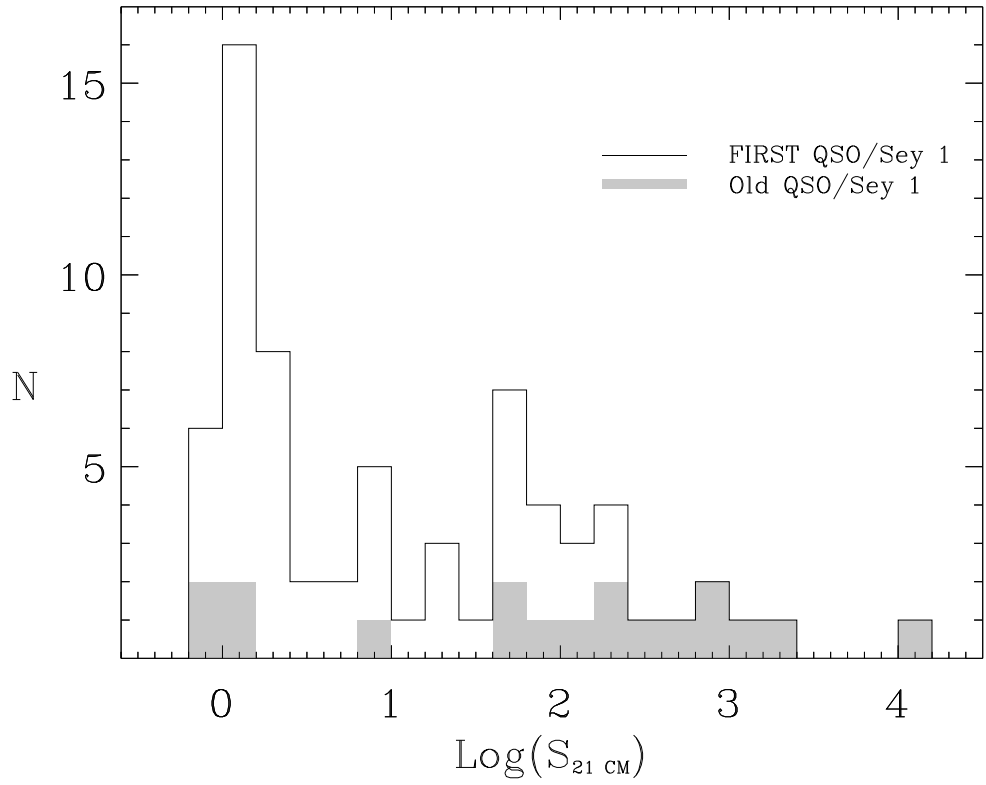}
\caption{Distribution of observed nuclear 21 cm flux for the QSO
sample.  Previously known QSOs are shaded.}
\end{figure}

\begin{figure}[t]
\plotone{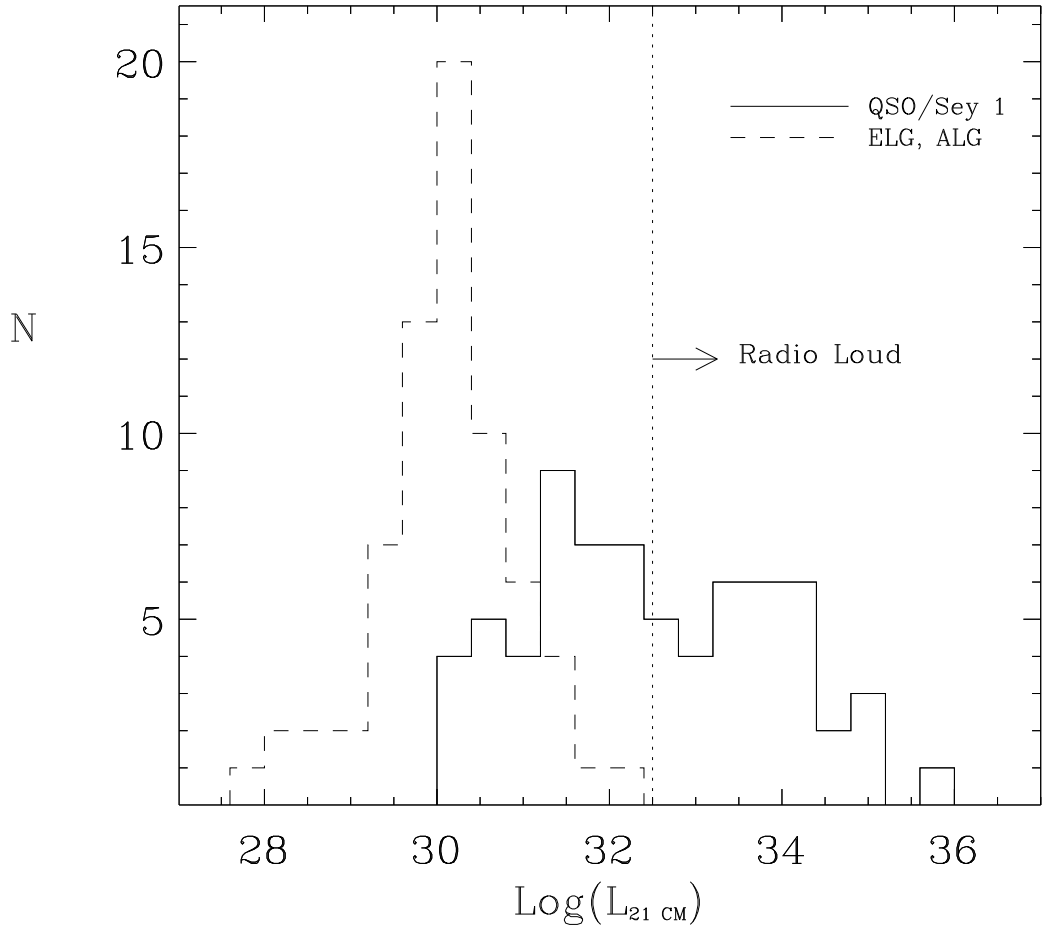}
\caption{Histogram of the QSO/Seyfert~1 nuclear radio
luminosities; there is a suggestion of a bimodal distribution between
radio loud and radio quiet.  The distribution of narrow emission line
and absorption line galaxy radio luminosities (dashed histogram) is
plotted for comparison.}
\end{figure}

\begin{figure}[t]
\plotone{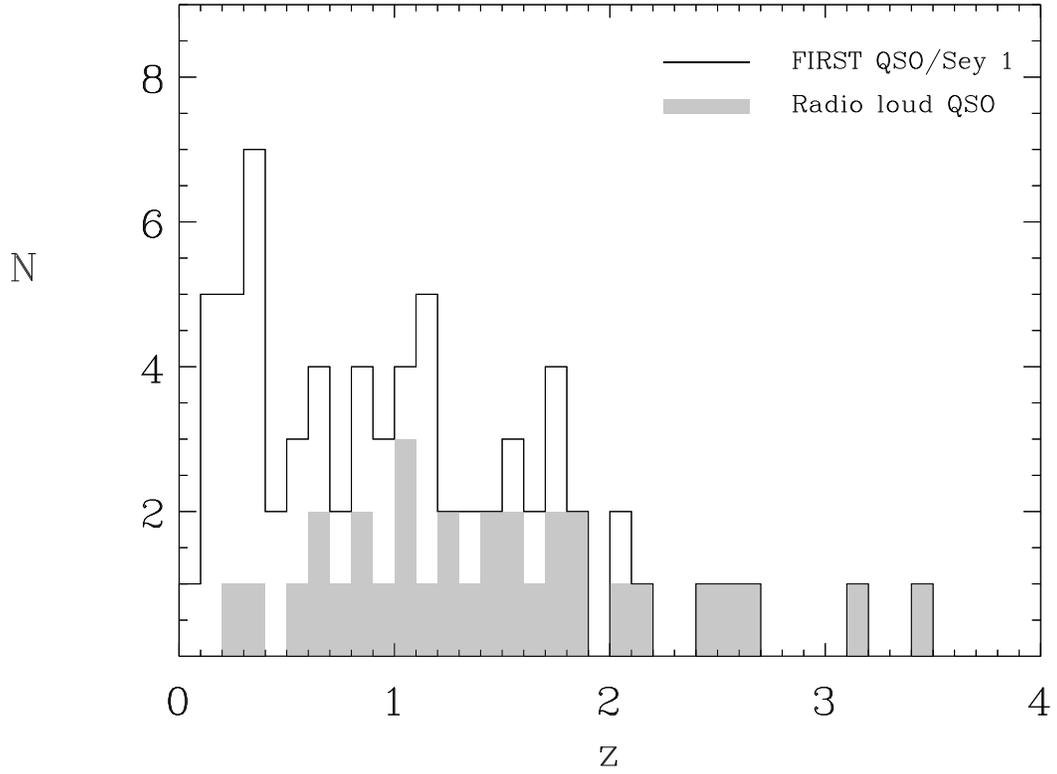}
\caption{Redshift distribution of the QSO sample.  Radio loud sources
are shaded.}
\end{figure}

\begin{figure}[t]
\plotone{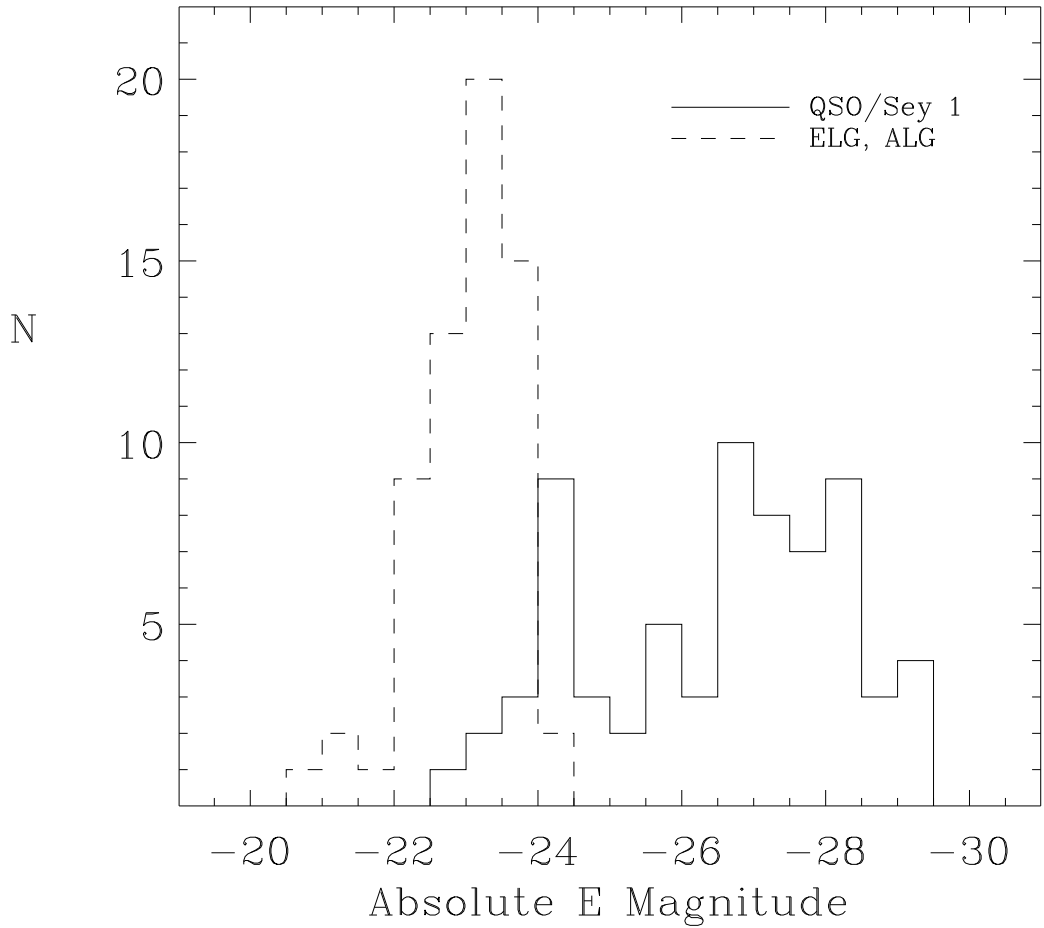}
\caption{Histogram of the QSO/Seyfert~1 optical (red) absolute
magnitudes (solid line); the excess in the faint tail can be explained
by the lower luminosity Seyfert~1 objects present in the sample.  The
distributions of narrow emission line and absorption line galaxies are
indistinguishable and are plotted as a single histogram (dashed line)
for comparison.}
\end{figure}

\begin{figure}[t]
\plotfiddle{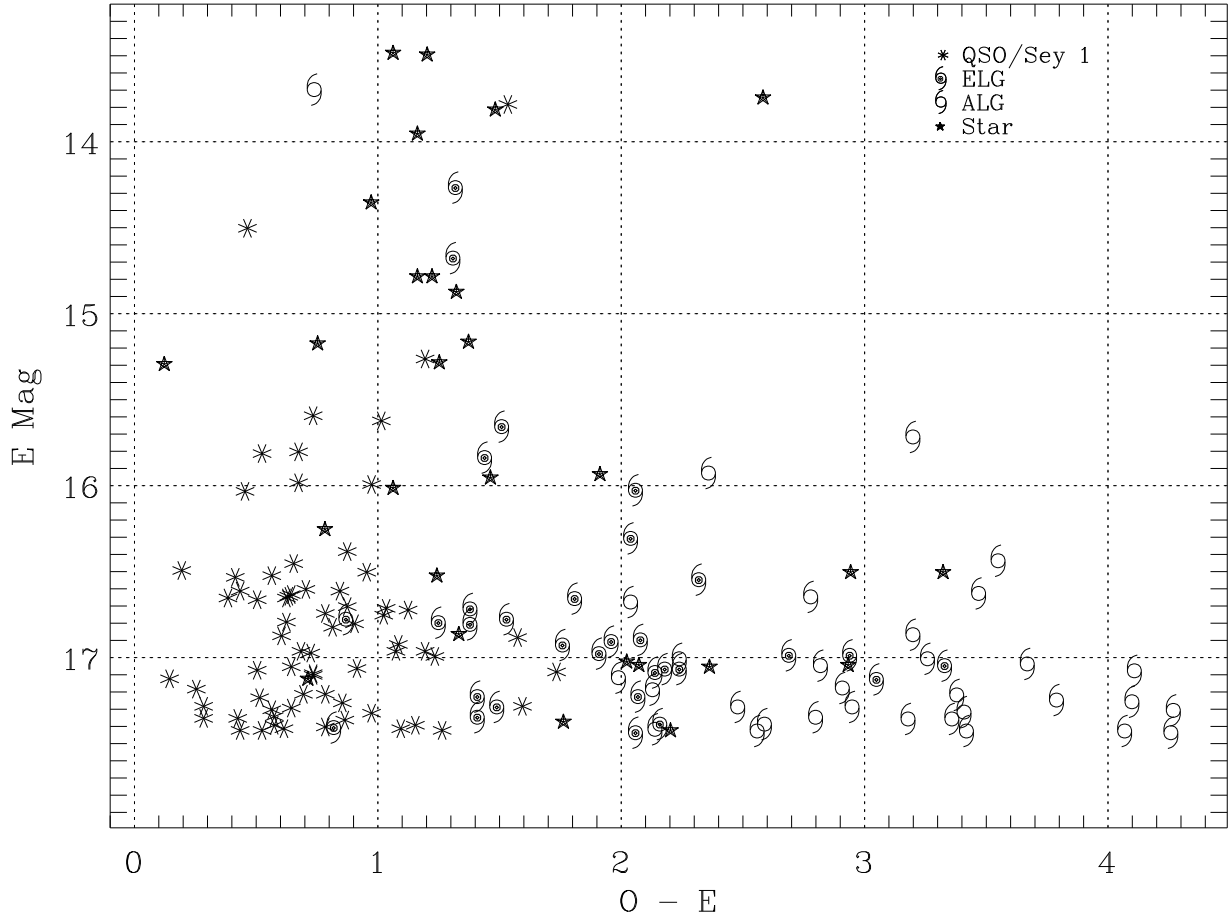}{5in}{90}{75}{75}{300}{0}
\caption{Color magnitude diagram of the entire spectroscopic sample.
The reddest QSO has O-E = 1.74; emission line galaxies have
colors intermediate between the QSOs and absorption line systems.}
\end{figure}

\end{document}